\documentclass[aps,prb,twocolumn,superscriptaddress, show pacs]{revtex4-1}
\usepackage{bm, amsmath}
\usepackage{graphicx}
\usepackage{color}
\usepackage{epstopdf}
\usepackage{graphicx}
\usepackage{dcolumn}
\usepackage{bm}
\usepackage{subfigure}

\begin{document}


\title{Skyrmion robustness  in non-centrosymmetric magnets with axial symmetry: The role of anisotropy and tilted magnetic fields}

\author{A. O. Leonov}
\thanks{leonov@hiroshima-u.ac.jp}
\affiliation{Center for Chiral Science, Hiroshima University, Higashi-Hiroshima, Hiroshima 739-8526, Japan}
\affiliation{Department of Chemistry, Faculty of Science, Hiroshima University Kagamiyama, Higashi Hiroshima, Hiroshima 739-8526, Japan}

\author{I. K\'ezsm\'arki}
\affiliation{Experimental Physics V, Center for Electronic Correlations and Magnetism, University
of Augsburg, Augsburg 86135, Germany}
\affiliation{Department of Physics, Budapest University of Technology and Economics and MTA-BME Lend\"ulet
Magneto-Optical Spectroscopy Research Group, Budapest 1111, Hungary}

\begin{abstract}
%
We investigate the stability of N\'eel skyrmions against tilted magnetic fields, in polar magnets with uniaxial anisotropy ranging from easy-plane to easy-axis type.
We construct the corresponding phase diagrams and investigate the internal structure of skewed skyrmions with displaced cores.
We find that moderate easy-plane anisotropy increases the stability
range of N\'eel skyrmions for fields along the symmetry axis, while
moderate easy-axis anisotropy enhances their robustness  against
tilted magnetic fields.
We stress that the direction, along which the skyrmion cores are
shifted, depends on the symmetry of the underlying crystal lattice.
The cores of N\'eel skyrmions, realized in polar magnets with C$_{nv}$ symmetry, are displaced either along or opposite to the off-axis (in-plane) component of the  magnetic field depending on the rotation sense of the magnetization, dictated by the sign of the Dzyaloshinskii constant.  
The core shift of antiskyrmions, present in chiral magnets with
D$_{2d}$ symmetry, depends on the in-plane orientation of the
magnetic field and can be parallel, anti-parallel, or perpendicular
to it.
%
%
%
We argue that the role of anisotropy in magnets with axially symmetric crystal structure is different from that in cubic helimagnets.
%
%
Our results can be applied to address recent experiments on polar magnets with C$_{3v}$ symmetry, GaV$_4$S$_8$ and GaV$_4$Se$_8$.
%
%

%
\end{abstract}

\maketitle

\section{The Introduction}
%
%
In magnetic compounds lacking inversion symmetry, the underlying crystal structure induces a specific asymmetric exchange coupling, the so-called Dzyaloshinskii-Moriya  interaction (DMI), which can stabilize long-period spatial modulations of the magnetization
with a fixed rotation sense \cite{Dz64,Bak}.
Within a continuum approximation for magnetic properties, the DMI is expressed by inhomogeneous invariants 
 involving first derivatives of the magnetization $\textbf{M}$ with respect to the spatial coordinates: 
\begin{equation}
\mathcal{L}^{(k)}_{i,j} = M_i \frac{\partial M_j}{\partial x_k} - M_j  \frac{\partial M_i}{\partial x_k}.
\label{Lifshitz}
\end{equation}
Depending on the crystal symmetry,  certain combinations of these Lifshitz invariants (LI)   can contribute to the magnetic energy of the material \cite{Bogdanov89}.

%
In the last few years, a renewed interest in non-centrosymmetric
magnets has been inspired by the discovery of two-dimensional
localized modulations, commonly called magnetic skyrmions
\cite{Bogdanov94,Yu10,Roessler11}.
Recently, skyrmion lattice states (SkL) \cite{Yu10,Yu11} and isolated skyrmions (IS) \cite{Romming13,Romming15} were discovered in bulk crystals of non-centrosymmetric magnets near the magnetic
ordering temperatures \cite{Muehlbauer09,Wilhelm11,Kezsmarki15} and in nanostructures with confined geometries over larger temperature
regions\cite{Yu10,Yu11,Du15,Liang15}.
The small size, topological protection and easy manipulation of skyrmions by electric fields and currents \cite{Schulz12,Jonietz10,Hsu17} generated enormous interest in their applications in information storage and processing \cite{Sampaio13,Tomasello14}.

\begin{figure}[tb]
\includegraphics[width=7.2cm]{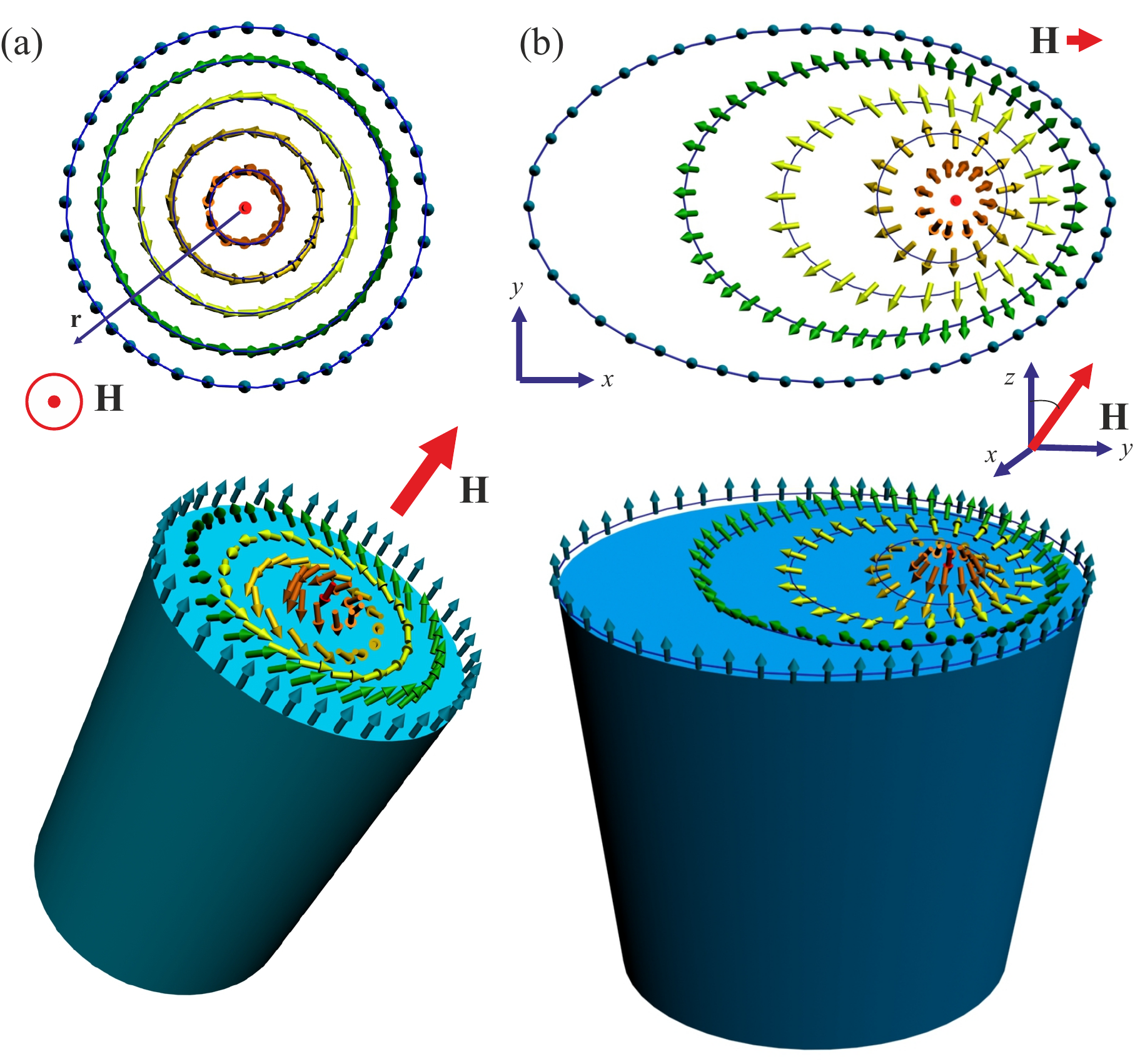}
\caption{
\label{Fig1} (Color online)  The spin texture of Bloch (a) and N\'eel (b) skyrmions in a tilted magnetic field. While the Bloch skyrmions in cubic helimagnets co-align their axes with the field, the N\'eel 
skyrmions displace their cores and lock their axes to the the high-symmetry (polar) axis of the host crystal.
}
\end{figure}

%
The 
DMI  provides not only a unique stabilization mechanism
\cite{Bogdanov94}, protecting magnetic skyrmions from radial
instability, but also governs their internal structure.
In Ref. \onlinecite{Bogdanov89} by Bogdanov and Yablonskii, three distinct types of skyrmions and two types of antiskyrmions (having opposite topological charge as compared with skyrmions) were predicted to occur in non-centrosymmetric magnets with different crystallographic symmetries.

\begin{figure}[tb]
\includegraphics[width=8.2cm]{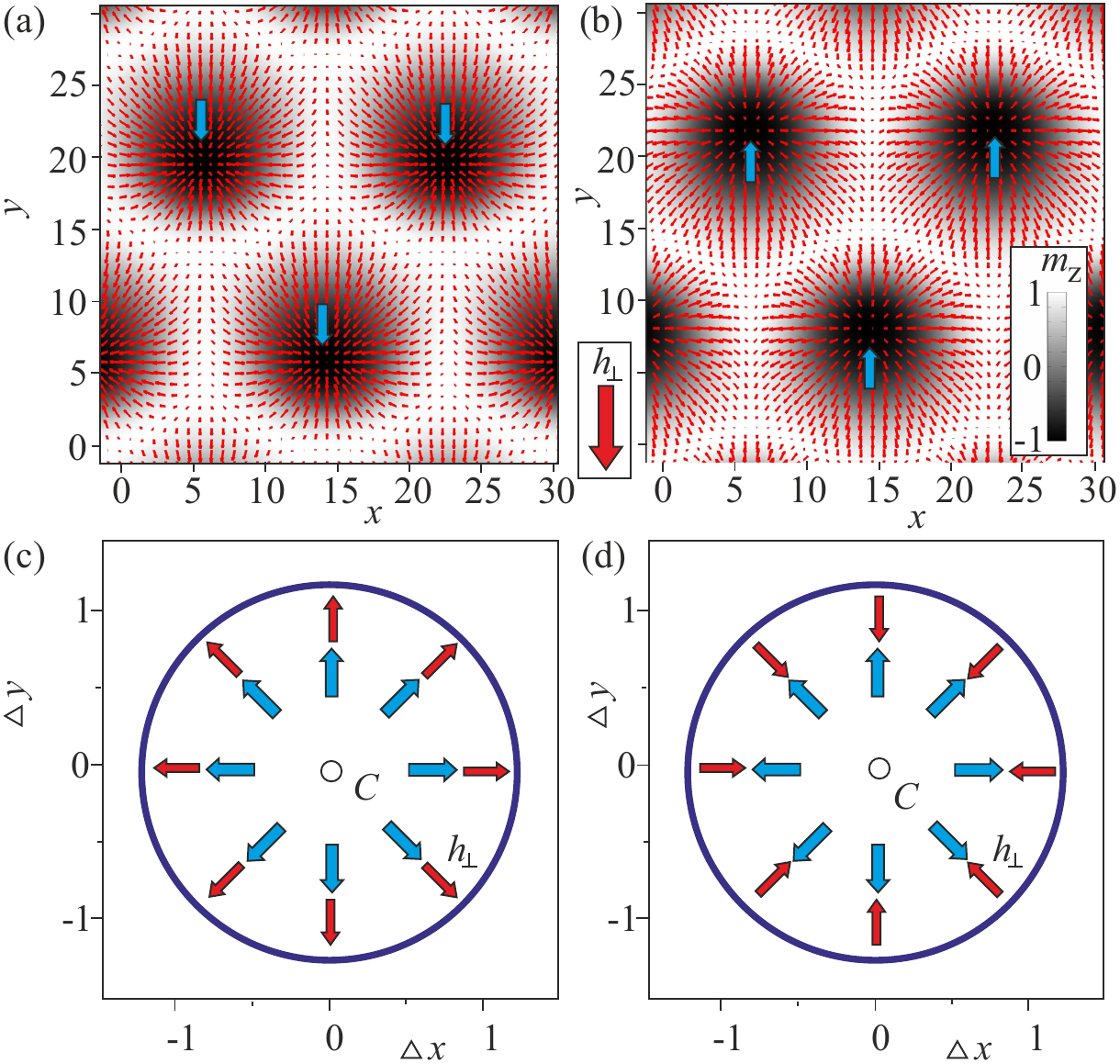}
\caption{
\label{structure} (Color online) N\'eel SkL in an oblique magnetic field in magnets with C$_{nv}$ symmetry, $h=0.2, \,\alpha=0.7$. (a), (b) the color plots of the $m_z$-component of the magnetization plotted for two types of N\'eel skyrmions realized with opposite signs of the Dzyaloshinskii constant $D$ in Eq. (\ref{functional}). In-plane components of the magnetization are shown with thin red arrows. The in-plane projection of the magnetic field $h_{\bot}$ is shown by the thick red arrow and is the same for both cases. $h_{\bot}$ leads either to the parallel (a) or antiparallel (b) shift of the skyrmion cores for the "in-ward" and "out-ward" rotation sense of the magnetization, respectively.
(c), (d) Displacement of skyrmion centers plotted with respect to the skyrmion centers in a "stright" magnetic field with $\alpha=0$, point $C$. The radius of the curve shows the magnitude of the core  shift. Blue and red arrows show corresponding directions of the field and core shift in each point of the curve.
}
\end{figure}

The first observed type of skyrmions is called \textit{Bloch} skyrmions, whose internal structure is schematically shown in Fig. \ref{Fig1} (a).
It is the most ubiquitous skyrmionic archetype observed in materials with B20 structure like MnSi \cite{Muehlbauer09}, FeGe \cite{Wilhelm11} or Cu$_2$OSeO$_3$ \cite{Seki12}.
The magnetization in Bloch skyrmions is perpendicular to the radial direction. 
Since the magnetic energy term corresponding to the DMI   is reduced to the isotropic form $ w_{DMI}=\mathbf{M}\cdot \mathrm{rot} \mathbf{M}$, the axes of Bloch skyrmions in cubic helimagnets always co-align with  the applied magnetic field, irrespective of the direction of the field. 
%


\textit{Antiskyrmions} were recently observed in Mn-Pt-Sn Heusler materials with acentric D$_{2d}$ crystal  symmetry \cite{Nayak}. 
They were shown to exist over a wide temperature interval both as IS and in ordered SkL. 
Although the antiskyrmions break the cylindrical symmetry and carry a quadrupolar moment of the magnetostatic charges,
they are invariant under all symmetry operations of D$_{2d}$, including  the $\overline{4}$ symmetry ($4$-fold rotation followed by inversion).
%

Polar magnets with C$_{nv}$ symmetry, such as  GaV$_4$S$_8$ \cite{Kezsmarki15} and GaV$_4$Se$_8$ \cite{Bordacs17,Fujima17}, 
host \textit{N\'eel} skyrmions in which the magnetization rotates radially  from the skyrmion center as shown in Fig. \ref{Fig1} (b).
The helicity angle $\chi=0$ ("out-ward" rotation of the magnetization) or $\chi=\pi$ ("in-ward" rotation of the magnetization) depends on the sign of the Dzyaloshinskii constant in model (\ref{functional}) , described later, and endows the N\'eel skyrmions with a magnetic monopole moment on the contrary to Bloch skyrmions with a magnetic toroidal moment \cite{Spaldin08}.
Moreover, N\'eel skyrmions are dressed with an electric polarization \cite{Ruff15,Wang15}   and induce internal magnetic charges due to the non-zero divergence of the rotating magnetization vector \cite{Bogdanov94}.
%
%

In the  C$_{nv}$ and D$_{2d}$  
symmetry classes investigated here, no Lifshitz invariants  along the high-symmetry axis, the $z$ direction, are  present \cite{Bogdanov89}.
Therefore, only modulated magnetic structures with wave vectors perpendicular to the $z$ axis are favoured by the DMI.
As it will be shown in this work, in magnetic fields tilted with respect to the high-symmetry axis both the N\'eel skyrmions  and antiskyrmions \cite{Nayak} shift their cores with respect to the in-plane component of the field instead of coaligning
 their axes with the field as it happens for Bloch skyrmions in cubic helimagnets,  the so-called locking of the skyrmion axes. 

LI (\ref{Lifshitz}) define a set of competing modulated phases in the  phase diagrams specific to the different crystallographic classes.
For the C$_{nv}$ and D$_{2d}$ 
symmetry classes, usually only spirals, skyrmions (or antiskyrmions), and collinear spin-structures are identified \cite{Nayak,Kezsmarki15}.
%
%
There is no competing conical state for fields  applied along the high-symmetry axis. 
Although a transverse conical state arises for fields spanning large angles with the high-symmetry axis \cite{Bordacs17}, but this is fundamentally different from the longitudinal conical state in cubic helimagnets, where the $\mathbf{q}$-vector co-aligns with the field.
The lack of the longitudinal conical state may lead to wide stability regions of skyrmions even in bulk materials with C$_{nv}$ and D$_{2d}$ 
symmetry classes.
In contrast in bulk cubic helimagnets,  the longitudinal conical phase is present for arbitrary direction of the magnetic field and restricts the region of skyrmion stability to a small pocket near the ordering temperatures (A-phase) \cite{Muehlbauer09,Wilhelm11}.

In the present paper, we investigate the robustness of N\'eel 
skyrmions with displaced cores in oblique magnetic fields and the role  
of the uniaxial anisotropy (UA).
Lacunar spinels GaV$_4$S$_8$ and GaV$_4$Se$_8$ \cite{Kezsmarki15,Ehlers17,Bordacs17} bear  uniaxial anisotropy of  easy-axis and  easy-plane type, respectively, as they undergo a Jahn-Teller transition from a room-temperature cubic structure into a low-temperature rhombohedral polar (C$_{3v}$) structure \cite{Hlinka16,Ruff15,Wang15}.
From the experimental phase diagrams \cite{Kezsmarki15,Fujima17,Bordacs17}, showing a temperature dependent ratio of $h_{c1}/h_{c2}$, where $h_{c1}$ is the field of the first-order phase transition between spirals and SkL and $h_{c2}$ is the field of SkL to ferromagnetic state transition, one can conclude that the magnitude of effective UA varies with temperature (see also Fig. \ref{correspondence}).
In particular, the magnetic phase diagrams of GaV$_4$S$_8$ obtained for different orientations of the magnetic field  (see e.g. Fig. 2 in Ref. \onlinecite{Kezsmarki15}) exhibit tri-critical points and by their topology resemble the theoretical phase diagrams constructed in Refs. \onlinecite{Butenko10,Leonov16a}.
Therefore in the present paper, we construct the phase diagrams (Fig. \ref{PD}) on the plane spanned by the axial and in-plane components of the magnetic field for both signs and various values of
 the UA. 
We show that the easy-axis anisotropy decreases the region of SkL stability and promotes the formation of ISs, whereas the easy-plane anisotropy enhances the SkL stability.
We also discuss the internal structure of modulated states and the way how the theoretical phase diagrams could be set up in correspondence with the experimental ones.
We also study the deformation of antiskyrmions in oblique magnetic fields, realized in Heusler compounds with tetragonal crystal structure.
This compound family  provide a perfect platform for the design of magnetic configurations with variable uniaxial anisotropy \cite{Nayak}.

\begin{figure}[tb]
\includegraphics[width=8cm]{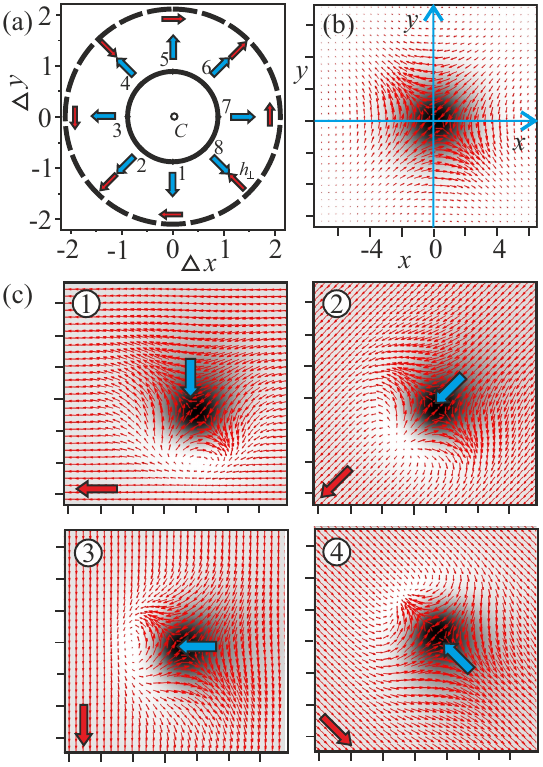}
\caption{
\label{structure2} {\color{black} (Color online) Isolated antiskyrmions in an oblique magnetic field plotted for magnets with D$_{2d}$ symmetry.
(a) Displacement of skyrmion centers plotted for $\alpha=0.7$ (solid line) and $\alpha=\pi/2$ (dashed line). The magnitude of the field is $h=0.6$ for both cases. The distance $\Delta$ by which the skyrmions shift their cores with respect to the point $C$ is tuned by the value of $\alpha$. Blue and red arrows in each point of the curve show the directions of the skyrmion shift and the  direction of the field, respectively.
(b) The spin structure of an isolated antiskyrmion in a field coaligned with its axis, the $\overline{4}$ axis of the host crystal. The color scale indicates $m_z$-component of the magnetization (the color scale is the same as in Fig. \ref{structure}). In-plane components of the magnetization are shown with thin red arrows in magnetic fields coaligned with its core. The $x$ and $y$ axes are selected in a way that $xz$ and $yz$ planes are the mirror planes of the D$_{2d}$ symmetry. Grating period of $x$ and $y$ axes is 2.  The point $C$  in (a) is the center of  an antiskyrmion.
(c)  The spin structure of isolated antiskyrmions with displaced cores.  The in-plane projection of the magnetic field $h_{\bot}$ is shown by thick red arrows. }
}
\end{figure}

\section{The Model}

The magnetic energy density of a non-centrosymmetric ferromagnet with C$_{nv}$ and  D$_{2d}$ symmetry can be written as the sum of the exchange, the DMI, Zeeman, and the anisotropy energy density contributions, correspondingly:
\begin{equation}
w=\sum_{i,j}(\partial_i m_j)^2+w_{DMI}-\mathbf{m}\cdot\mathbf{h}-k_u m_z^2.
\label{functional}
\end{equation}
where spatial coordinates $\mathbf{x}$ are measured in units of the characteristic length of modulated states $L=A/D$. $A>0$ is the exchange stiffness, $D$ is the Dzyaloshinskii constant.
%
The unit vector along the magnetization $\mathbf{M}$ is $\mathbf{m} = \mathbf{M}/|\mathbf{M}|$.
$\mathbf{h} = \mathbf{H}/H_0$ is the applied magnetic field, $H_0 = D^2/A|\mathbf{M}|$.
The magnetic field  has a tilt angle $\alpha$ with the $z$ axis, i.e. 
\begin{equation}
h_z=h\cos\alpha, \, h_{\bot}=h\sin\alpha.
\label{field}
\end{equation}
%
%
The UA $k_u=K_uM^2A/D^2$. In the case of N\'eel skyrmions, $k_u$  is the effective uniaxial anisotropy which includes the intrinsic uniaxial anisotropy ($K_u$) and the anisotropy due to the stray field energy \cite{Bogdanov94}. 

%
The DMI energy density  has the following form:
\begin{align}
&C_{nv}:  m_x\partial_x m_z-m_z\partial_x m_x+m_y\partial_y m_z-m_z\partial_y m_y, \nonumber\\ 
&D_{2d}:  m_x\partial_y m_z-m_z\partial_y m_x+m_y\partial_x m_z-m_z\partial_x m_y
\end{align}
where $\partial_x=\partial/\partial x,\, \partial_y=\partial/\partial y$.
%
In both cases, only modulated phases with the propagation directions perpendicular to the polar axis of C$_{nv}$ or to the tetragonal axis in magnets with D$_{2d}$ symmetry, i.e. in the $xy$-plane of model (\ref{functional}), are favored.

%
Functional (\ref{functional}) includes only the main energy contributions essential to stabilize modulated states and an additional UA. 
As the main goal of the paper is to investigate the role of UA in the stability regions of skyrmions with shifted cores, we do not take into account other anisotropic contributions.
In particular, besides the single-ion type anisotropy, considered here, we neglect
 the difference in exchange couplings for two sets of bonds  present in  lacunar spinels GaV$_4$Se$_8$ \cite{Bordacs17} and GaV$_4$S$_8$ \cite{Kezsmarki15} as well as the exchange anisotropy. 
Moreover, we do not capture the influence of thermal fluctuations near the ordering temperatures.

\begin{figure}[tb]
\includegraphics[width=6cm]{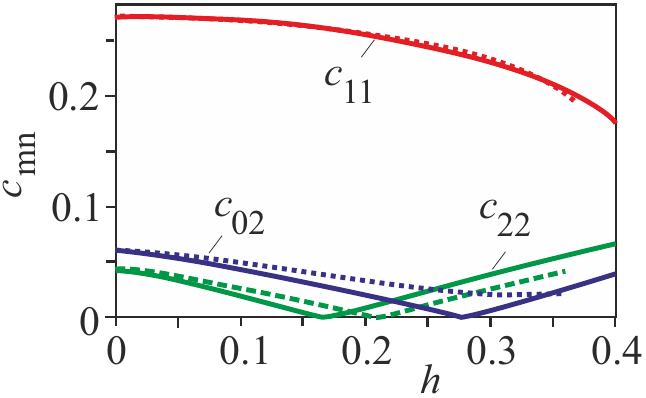}
\caption{
\label{fourier2} (Color online) Coefficients $c_{11}$ (red line), $c_{02}$ (blue line), and $c_{22}$ (green line) of the fast Fourier transform (\ref{fourier})  for "stright" skyrmions with $\alpha=0$ (solid lines) and elongated skyrmions with $\alpha=0.9$ (dotted lines) See text for details. $c_{11}$ is the amplitude of the fundamental harmonics (six-spot pattern characteristic to SkL),
while $c_{02}$ and $c_{22}$ are the amplitudes of higher-order harmonics. 
}
\end{figure}

\section{Results and discussion}

\subsection{Chiral modulations in C$_{nv}$ magnets}

Three magnetic phases obtained as the solutions of the model (\ref{functional}) can be identified as follows:

(i) \textit{ Spirals.}  For the case of C$_{nv}$ magnets the spirals represent cycloids with the rotation plane of the magnetization containing the wave vector $\mathbf{q}$ and the high-symmetry ($C_n$) axis.
For D$_{2d}$ crystals, the type of the spiral depends on the in-plane orientation of the $\mathbf{q}$-vector: it is a helicoid (a Bloch spiral) for $\mathbf{q}\,||\,x$ and $\mathbf{q}\,||\,y$, while cycloids are formed
 for diagonal directions as can be discerned in the corresponding structure of the antiskyrmion in Fig. \ref{structure2}.  The $x$ and $y$ axes are selected in a way that $xz$ and $yz$ planes are the mirror planes of the D$_{2d}$ symmetry.

\begin{figure}[tb]
\includegraphics[width=6.2cm]{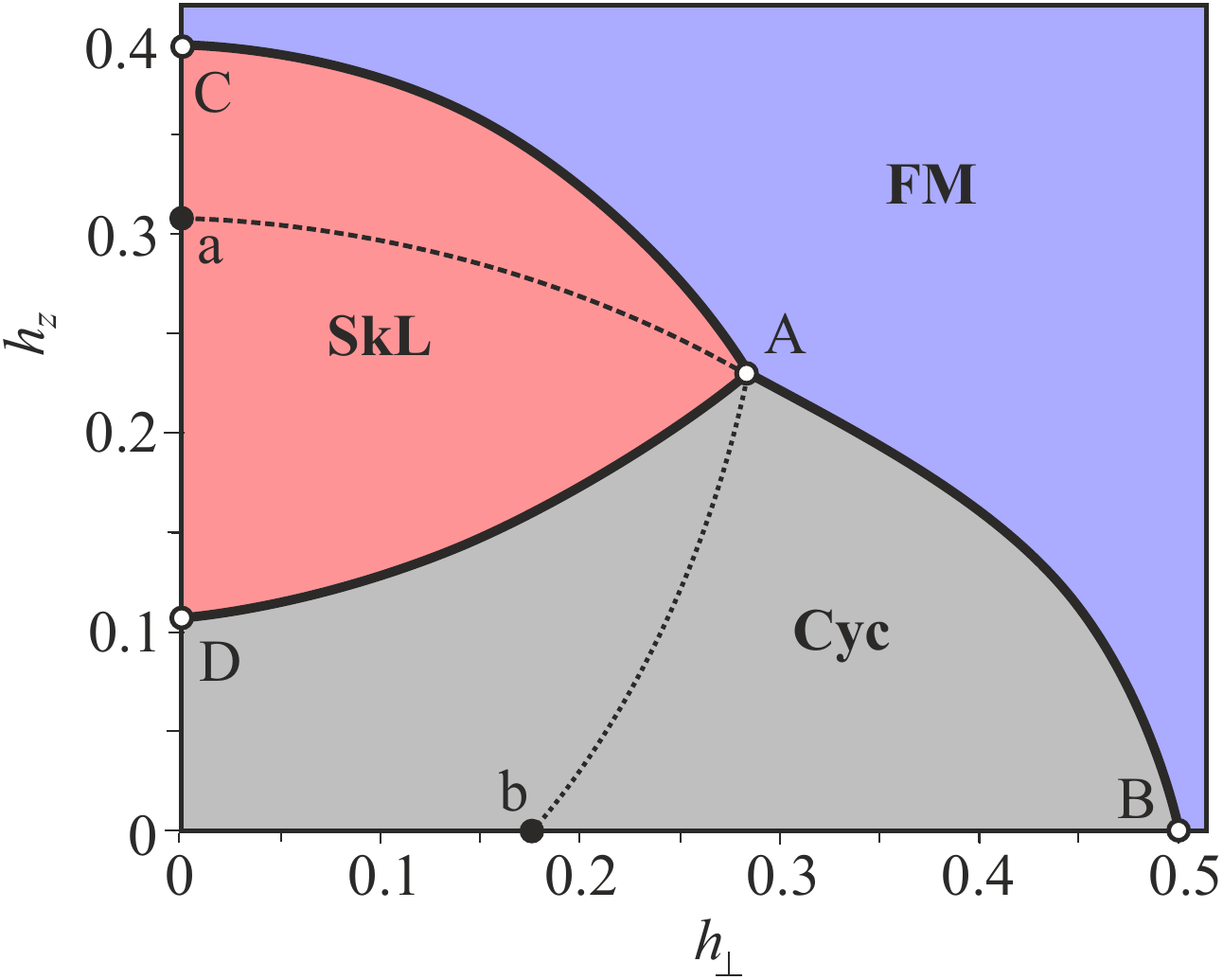}
\caption{
\label{PD0} (Color online) {\color{black} The phase diagram 
for model (\ref{functional}) plotted for zero UA $k_u=0$ on the plane $(h_{\bot},h_z)$ of the field components. Filled areas indicate the regions of global stability for the SkL (red) and spirals (gray).  In the saturated state (blue area), skyrmions exist as isolated objects.
Thin dashed line $a-A$  designates the phase transition between the metastable spiral state and the polarized ferromagnetic phase.
Thin dotted line $b-A$ is the line of the elliptical instability of the metastable SkL with respect to the spiral state (see text for details). }
}
\end{figure}

\begin{figure*}[tb]
\includegraphics[width=17.2cm]{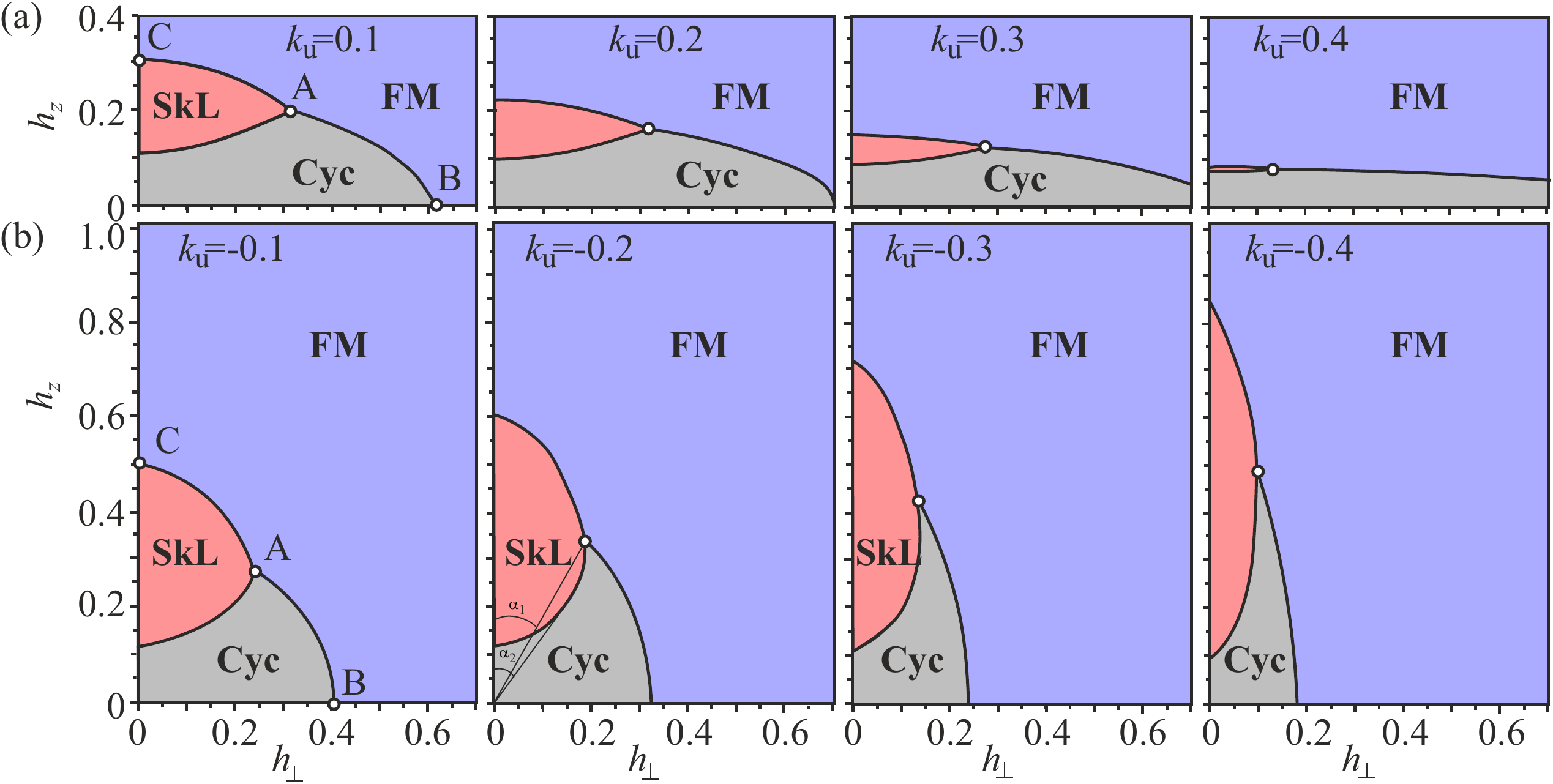}
\caption{
\label{PD} (Color online)  Magnetic phase diagrams as obtained for model (\ref{functional}) on the $h_{\perp}-h_z$ plane  of the field components with the variable constant of the uniaxial anisotropy ($k_u$) for the easy-axis (a) and easy-plane (b) cases. $h_{\perp}$ and $h_z$ stand for the magnetic field components perpendicular and along the high-symmetry (C$_n$) axis. {\color {black} The second graph in (b) also shows how the critical angles $\alpha_1$ and $\alpha_2$ signifying the onset of the reentrant phenomenon are defined. In the interval  $\alpha_1-\alpha_2$, with increasing field
the SkL transforms back to a skewed spiral state via the first-order phase transition. }
}
\end{figure*}

(ii) \textit{ Skyrmions.}  Skyrmions can exist either in the form of a thermodynamically stable SkL (the corresponding regions are red shaded in the phase diagrams of Figs. \ref{PD0} and \ref{PD})  or as metastable IS within the stable homogeneous state (the corresponding regions are blue-shaded in Figs. \ref{PD0} and \ref{PD}). Isolated N\'eel skyrmions in bulk systems with C$_{nv}$ have not been experimentally observed yet, although they were reported in thin films with interface-induced DMI \cite{Romming13,Leonov16a}. Isolated antiskyrmions in bulk Heusler alloys were  reported in Ref. \onlinecite{Nayak}.

When the magnetic field is parallel with the  high-symmetry (C$_n$) axis, i.e. $h_{\perp}=0$, the period of the cycloid and its anharmonicity is gradually increased with increasing field. When the magnetic field is perpendicular to the  high-symmetry axis, i.e. $h_{z}=0$, a transverse conical state emerges, with a uniform magnetization developing parallel to the field, besides the magnetization component rotating in the plane perpendicular to the field. The opening angle of the cone is decreased with increasing field.
As the field points in a general oblique angle, the modulated phases are skewed.
In particular, skewed cycloids with the $\textbf{q}$-vectors along $y$ axis, when $h_{\perp}$ is along the $x$ axis,  can be expanded by
cycloids with magnetization rotating in the $yz$ plane (period of such a cycloid gradually increases with increasing $h_z$-component of the field), plus a transverse component oscillating in the $xy$ plane and a uniform component.

%

(iii) \textit{ Angular homogeneous phases.}  The magnetization of these field polarized ferromagnetic phases is tilted away from the $z$ axis in the presence of oblique fields. 

\subsection{The internal structure of skewed skyrmions}

(i) \textit{ Skewed N\'eel skyrmions}.
The shape of the N\'eel skyrmions in tilted magnetic fields, as shown in Fig. 3, deforms in the following way: the cores of the skyrmions are shifted from the center of the unit cell of the SkL, but the lattice retains the stability against transformation into spirals.
Whether the skyrmion centers shift along or opposite to $h_{\bot}$ depends on  the sign of the Dzyaloshinskii constant $D$ in Eq. (\ref{functional}): N\'eel skyrmions with the "in-ward" sense of the magnetization rotation (Fig. \ref{structure} (a)) have a  core displacement along $h_{\bot}$, whereas  N\'eel skyrmions with the "out-ward" magnetization rotation (Fig. \ref{structure} (b)) have a core shift opposite to $h_{\bot}$.
The effect is easily understood, as the in-plane component of the magnetic field $h_{\bot}$ increases the area, where the magnetization points along the field. 
The corresponding shift of skyrmion cores is marked by blue arrows in Fig. \ref{structure} (a) and (b).

Figs. \ref{structure} (c) and (d) show the displacement of skyrmion centers as a function of the orientation of the in-plane field component for $h=0.2,\,\alpha=0.7$. 
The values $\Delta x$ and $\Delta y$, the components of the core shift along the two axis, are
relatively small as compared with the period  of a spiral in zero magnetic field.
The displacement of the core is of the same magnitude, irrespective of the orientation of the in-plane component of the field and does not reflect the six-fold symmetry of the SkL unit cell.
%

In Ref. \onlinecite{Lin15}, it was shown that the skewed skyrmions have anisotropic inter-skyrmion interaction, i.e. the interaction energy between two skyrmions depends on their relative orientation and leads to the distorted triangular lattice.
Nevertheless, the skyrmion-skyrmion interaction remains repulsive with the strongest repulsion between skyrmions in the direction where they are elongated \cite{Lin15}.
This phenomenon singles out the  N\'eel 
skyrmions from the Bloch skyrmions in B20 compounds:
while Boch skyrmions orient their cores along the magnetization of the ferromagnetic host phase, skyrmions in magnets with axially symmetric crystal structure experience an orientational confinement, with their cores nearly fixed to the direction of the high-symmetry axis.
%
%
%
%
%
Therefore for B20 materials, it is instructive first to make a full analysis of the magnetization processes for homogeneous states \cite{Leonov08} and to define the global and local minima of the energy functional (\ref{functional}).
Then, the stable (metastable) SkL will orient their axes along the corresponding homogeneous states and form multiple skyrmionic domains with non-trivial topological boundaries between them.

In thin films of B20 materials, the reorientation process of the Bloch skyrmions in the tilted fields is a more delicate issue, leading to an undulated shape of the skyrmion cylinders \cite{Wang17}.
Such a skyrmion deformation must minimize, in particular, contributions from so called surface twists,  modifying the structure of 
skyrmions near the surface and even leading to their thermodynamical stability (see Ref. \onlinecite{Leonov16b} for details),  and from the demagnetization effects.
In Ref. \onlinecite{Yokouchi14} the process of skyrmion core rotation in tilted magnetic fields has been observed in thin epitaxial films of MnSi.
The increase of  the angle $\alpha$ results in the effective increase of the film thickness for skyrmions and, thus, in  the appearance of the conical phase which is stable for thicker MnSi-films (in accordance with the theoretical phase diagram constructed in Ref. \onlinecite{Leonov16b}).

%
%
%
%

%

(ii)  \textit{Skewed antiskyrmions}.
In case of the antiskyrmions the core displacement direction depends on the orientation of the in-plane field  component $h_{\bot}$ with respect to the coordinate axes.
%
The modulations propagating parallel to these axes are helical, while the modulations propagating along the $\pm$45 degree lines are cycloidal (Fig. \ref{structure2} (b)).
%
%
The core shift is perpendicular to the field if $h_{\bot}$ is coaligned with the coordinate axes, i.e. for $h_x=0, h_y=h_{\bot}$ (or $h_x=h_{\bot}, h_y=0$), corresponding to points 1,3,5,7 in Fig. \ref{structure2} (a) and (c).
The core shift occurs along and opposite to the field in points 2, 6 and 4,8 of Fig. \ref{structure2} (a) and (c), respectively. 
%

The distance $\Delta$, by which the skyrmion centers are displaced from their original position (point $C$) in the oblique magnetic field, depends on the value of the field and the angle $\alpha$: Fig. \ref{structure2} (a) shows two curves for $\alpha=0.7$ (solid line) and $\alpha=\pi/2$ (dashed line), with $h=0.6$ in both cases..


%

%

\subsection{Fourier components of skewed SkL}

The internal structure of the skyrmion lattice can also be characterized by the  behavior of the Fourier components in oblique magnetic fields.
Fig. \ref{fourier2} shows field dependence of the coefficients $c_{11}$ (red line), $c_{02}$ (blue line), and $c_{22}$ (green line) for "stright" skyrmions with $\alpha=0$ (solid lines) and elongated skyrmions with $\alpha=0.9$ (dotted lines). 
The Fourier components are introduced according to
\begin{align}
&m_z(x,y)=\sum_{mn} c_{mn}e^{i(mx+ny)}, \nonumber\\
&c_{mn}=\frac{1}{S}\int\int m_z(x,y) e^{i\,(mx+ny)}dxdy
\label{fourier}
\end{align}
where $S$ is the surface area of the corresponding unit cell for SkL.
%
The "stright" and elongated skyrmions can be distinguished by the behavior of their $c_{02}$ coefficients: for $\alpha=0.9$ these coefficients never become zero as is the case for $\alpha=0$.

The coefficients $c_{22}$ and $c_{02}$ for "stright" skyrmions become zero for some values of the field  and are small in the field region around $h=0.2$. 
In Ref. \onlinecite{Wilson14} (see Fig. 10\cite{Wilson14}), it was shown that for the case of B20 magnets at the field $h=0.2$ the difference between the energy densities of the SkL and the conical phase is minimal.
On this ground, it was suggested that the skyrmion lattice is stabilized with respect to the cones exactly around this field value and underlies the phenomenon of A-phase \cite{Muehlbauer09}.
Analysis of the Fourier components (\ref{fourier}) may justify the use of the triple-\textit{q} ansatz in this field interval \cite{Muehlbauer09}. 
Near the transition into the ferromagnetic state, however, the values of higher harmonics become significant and the ansatz in the form of the triple-\textit{q} becomes incorrect.

\subsection{Phase diagrams of states}

{\color{black}

By comparing the equilibrium energies of spirals, SkL and polarized FM states, we construct the phase diagrams  on the planes ($h_{\bot},h_z$) of the field components for different values of the uniaxial anisotropy $k_u$.
The areas of  magnetic states corresponding to the global minimum of the energy functional are indicated by different colors.

We start the  analysis of the phase diagrams with the $k_u=0$ case.
The SkL is thermodynamically stable within a curvilinear triangle $A-C-D$ (red-shaded region) with vertices $\textrm{A}(0.29,0.23)$, $\textrm{D}(0,0.11)$, and $\textrm{C}(0,0.4)$ \cite{Butenko10,Bogdanov94}.
The line $A-D$ is the line of the first-order phase transition between spirals and SkL.
At the line $A-C$ the SkL transforms into the polarized FM state whereas at the line $b-A$ (thin dotted line) the metastable SkL undergoes an elliptical  instability toward the spiral state (an elliptical instability of ISs was considered in Refs. \onlinecite{Bogdanov99,Leonov16a}).
Therefore, the point $A$ is a cusp point in which different lines characterizing different processes joint.
The point A also corresponds to the threshold angle $\alpha_{max}$  of an oblique magnetic field, where the thermodynamically stable SkL state still emerges. 
As a metastable solution, SkL exists even in a perpendicular magnetic field in the  range $0-b$ of Fig. \ref{PD0}.
For $\alpha>\alpha_{max}$ only skewed spirals (gray-shaded region) can be observed culminating in a conical phase at the point B ($\textrm{B}(0.5,0)$ for $k_u=0$).
The analysis of the type of the phase transition between skewed spirals and FM states as observed along the line $a-A-B$ has been done in Ref. \onlinecite{Laliena16}.

}

\begin{figure}[tb]
\includegraphics[width=8.2cm]{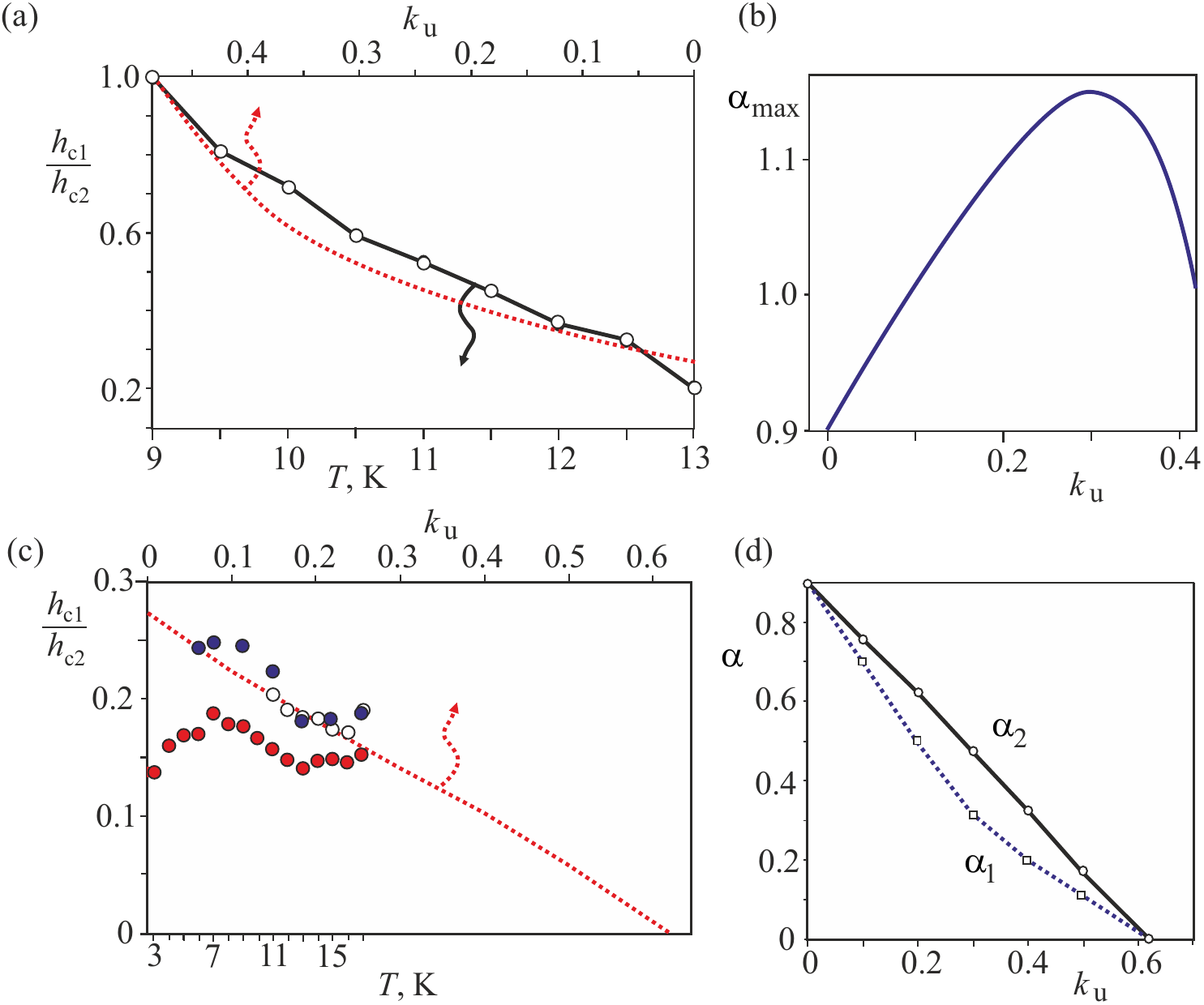}
\caption{
\label{correspondence} (Color online) (a) the temperature-dependent (lower scale, black line with circular markers) ratio $h_{c1}/h_{c2}$ as obtained from the experimental phase diagrams in Ref. \onlinecite{Kezsmarki15} and plotted together with the UA-dependent ratio obtained theoretically from model (\ref{functional}) (upper scale, red dotted line). (b) the maximal angle $\alpha_{max}$ of the tilted magnetic field corresponding to the point A at the phase diagram of Fig. \ref{PD} (a) and plotted in dependence on the constant $k_u$ of UA. For $\alpha$ exceeding this value, no SkL can be realized. (c) the  ratio $h_{c1}/h_{c2}$ plotted for the easy-plane UA: open circular markers show the values from Ref. \onlinecite{Bordacs17}, blue and red circular markers - from Ref. \onlinecite{Fujima17}. (d) $k_u$-dependence of the angles $\alpha_1$ and $\alpha_2$ signifying the onset and ending of the reentrant phenomenon at which the spiral state may appear from the skyrmion lattice at the increasing magnetic field.
}
\end{figure}

%
For \textit{the easy-axis} uniaxial anisotropy ($k_u>0$ in Eq. (\ref{functional}), Fig. \ref{PD} (a)), the existence range of the skyrmion lattice is the largest for the field co-aligned with the magnetic easy axis (point C in Fig. \ref{PD}) and decreases with  increasing tilting angle $\alpha$.
The angle $\alpha_{max}$ is a function of the uniaxial anisotropy $k_u$ as shown in Fig. \ref{correspondence} (b). 
Since the skyrmion cores are aligned along the easy axes, therefore, the stronger the anisotropy is, the higher  field is needed to shift the skyrmionic cores and the larger $\alpha_{max}$ can be achieved.
At the same time,  the UA suppresses the modulated phases.
Thus, the temperature or UA-dependence of $\alpha_{max}$ shows a maximum at $k_u=0.3$, which is realized at T=11K in GaV$_4$S$_8$. 

In Fig. \ref{correspondence} (a) for $\alpha=0$, we plot the ratio $h_{c1}/h_{c2}$:  a red dotted line  has been obtained from the theoretical phase diagram in Ref. \onlinecite{Butenko10} and has been plotted as a function of $k_u$ (corresponding to upper scale), while the solid black line (with circular markers) is the temperature dependence of $h_{c1}/h_{c2}$ as
extracted from the experimental data of Refs. \onlinecite{Kezsmarki15,Bordacs17}  for GaV$_4$S$_8$ and GaV$_4$Se$_8$ in $H\parallel [111]$ (lower scale).
By this, we make a correspondence between the experimental phase diagrams obtained for lacunar spinels GaV$_4$S$_8$ and GaV$_4$Se$_8$ and the theoretical phase diagrams in Fig. \ref{PD}.
Value $h_{c1}/h_{c2}=1$ in Fig. \ref{correspondence} (a) corresponds to the three-critical point with $k_u^{cr}=0.475$ and temperature $T^{cr}=9$K for GaV$_4$S$_8$\cite{Kezsmarki15}.
For $T<T^{cr}$, skyrmions are suppressed and only cycloids are observed.
For $T<5$ K ($k_u>0.617$) only the ferromagnetic state is present in the phase diagram. 
Thus the increasing  UA first suppresses the SkL and eventually the cycloids.

%

In GaV$_4$S$_8$ \cite{Kezsmarki15} at the structural transition, the lattice is stretched along one of the four $<111>$ body diagonals.
The four distortion directions correspond to the magnetic easy axes of the four structural domains coexisting below 44 K.
The magnetic state in each domain depends on the strength and orientation of the magnetic field with respect to the corresponding easy axis.
In Ref. \onlinecite{Kezsmarki15} the magnetic field was applied along the following directions in the cubic setting:
 $<100>\,(\alpha=0.955)$, $<110>\,(\alpha=0.615,\,\alpha=\pi/2)$, and $<111>\, (\alpha=0,\,\alpha=1.23)$.
The experimentally observed $\alpha_{max}=1.23$ is slightly larger than the theoretical value in Fig. \ref{correspondence} (b) which could be explained, e.g., by the  exchange anisotropy contribution or the slight rotation of N\'eel skyrmions neglected in our theory.

For \textit{the easy-plane} anisotropy ($k_u<0$ in Eq. (\ref{functional}), Fig. \ref{PD} (b)), the region of the skyrmion stability is significantly enhanced for $\alpha=0$.
Indeed, it was reported recently in GaV$_4$Se$_8$ \cite{Bordacs17,Fujima17} that SkL remains stable down to zero Kelvin.
This corroborates with the prediction by Randeria and co-workers as applied for SkL in polar materials with Rashba-type spin-orbit interaction \cite{Rowland16}.

The angular stability of SkL, however, impairs, as the $\alpha_{max}$ decreases with $k_u$ as discerned in Fig. \ref{correspondence} (d).
This is accompanied by the re-entrance phenomenon in some interval of angles $\alpha$: two successive first-order phase transitions from the spiral state to SkL and back take place with increasing  magnetic field.
In Fig. \ref{PD} (b) $\alpha_1$ corresponds to the point A and signifies the onset of the re-entrant spiral state, $\alpha_2$ coincides with the farthest point of the SkL stability region and signifies the limiting angle of the tilted field after which the re-entrance phenomenon disappears.
According to Fig. \ref{correspondence} (d) the largest interval $\Delta\alpha=\alpha_2-\alpha_1$  is reached for $k_u=0.3$.
No experimental observation of the predicted re-entrance phenomenon has been reported yet.

The values of $h_{c1}/h_{c2}$ extracted from the experimental data on GaV$_4$Se$_8$ in Ref. \onlinecite{Bordacs17} (white circles in Fig. \ref{correspondence} (c)) give the values of $k_u$ in the diapason $0.15-0.25$.
Only the temperature range $11-17$K was taken into account since for $T<11K$ some additional unknown phases appear which complicates the assignment of $h_{c1}$ and $h_{c2}$. 
These values slightly differ from the values extracted from Ref. \onlinecite{Fujima17} (blue and red circles in Fig. \ref{correspondence} (c)).
No new phases below $T<11$K have been reported in Ref. \onlinecite{Fujima17}.

The structure of IS surrounded by the canted ferromagnetic phase (blue-shaded regions in Figs. \ref{PD0} and \ref{PD}) was recently studied in Ref. \onlinecite{Leonov17}.
Such IS has asymmetric magnetic structure and exhibit anisotropic inter-skyrmion potential due to the intricate domain-wall region, which connects the core of the IS with the embedding canted ferromagnetic phase.
In Ref. \onlinecite{Lin15} the shape of isolated skyrmions has been investigated for the field $h\approx 0.4$ and $k_u=0$ in the units of Eq. (\ref{functional}), i.e., at the boundary between the skyrmion lattice and the homogeneous phase. With the increasing angle $\alpha$, as seen from the phase diagram, one gets onto the region of the stable skewed spiral. Therefore, a clear coexistence phase of the isolated skyrmion and the skewed spiral can be reached. 
For $h>0.5$ even for $\alpha=\pi/2$ ISs exist as localized entities within the homogeneously magnetized in-plane phase as also demonstrated for antiskyrmions in Fig. \ref{structure2} (a).

The theoretical phase diagrams in Fig. \ref{PD} 
reproduce the main aspects of the experimental phase
diagrams in Refs. \onlinecite{Kezsmarki15,Bordacs17} for both GaV$_4$Se$_8$ and GaV$_4$S$_8$ as compared with the isotropic case: i) the ratio of the critical fields required to reach the ferromagnetic state along and perpendicular to the polar c axis (i.e. the ratio $h_C/h_B$), ii) the extended or reduced angular stability range of SkL for the tilted fields in case of UA of easy-axis or easy-plane type,
 iii) the suppression or elongation of the stability range of the cycloidal states  against fields perpendicular to the polar axis (point B).
In particular, the value of the UA $k_u=0.25$ reproduces well the experimentally constructed phase diagram in Ref. \onlinecite{Bordacs17} and was chosen to match the ratio $h_C/h_B$ in Fig. \ref{PD} (b) and the temperature value in Fig. \ref{correspondence} (c).

%
%


We therefore argue that the role played by the UA is different in chiral B20 magnets and in considered materials with C$_{nv}$ and D$_{2d}$ symmetry.
In the first case, the UA of the easy-axis type (introduced by the axial deformation of the originally cubic structure) is invoked to suppress the conical phase \cite{Butenko10}.
SkL becomes thermodynamically stable in some interval of anisotropy parameters although in the rest of the phase diagram SkL remains a metastable state.
For the UA of the easy-plane type, only the conical phase is the global minimum of the functional (\ref{functional}).
To stabilize SkL in this case, one should find corresponding directions of an applied magnetic field.
In particular, cigar-like skyrmions can be stabilized with their axes lying in the easy plane together with the co-aligned magnetic field \cite{Wilson14}.

In the case of C$_{nv}$ and D$_{2d}$ acentric magnets, the UA modifies correspondingly the regions of the SkL stability promoting either ISs (easy-axis type) or enhancing the stability range of SkL (easy-plane type).
%
%
The easy-plane anisotropy may also lead to a stability of non-axisymmetric IS with unique properties \cite{Leonov17} as compared to ordinary axisymmetric skyrmions \cite{Leonov16a}.
%
%


%
\section{Conclusions}

%
%
In conclusion, we present theoretical studies on the robustness of skyrmions in tilted magnetic fields with special attention on the role of uniaxial magnetic anisotropy, which turns out to be a key factor governing the stability range of the modulated states. 
%
%
Together with earlier complementary studies on the cubic chiral magnets, our results on axially symmetric non-centrosymmetric magnets establish a consistent picture about the role of uniaxial anisotropy in the stability of modulated phases including spiral, skyrmion and antiskyrmion lattices.
%
Our findings can be tested experimentally in non-centrosymmetric magnets lacking the Lifshitz invariants along $z$, e. g. with C$_{nv}$, D$_{2d}$, and S$_4$ crystallographic symmetry.
We also described the deformations of skyrmions and antiskyrmions, both in lattices and individual ones, when the magnetic field is tilted away from this unique axis, which is the high-symmetry axis in these crystallographic classes.

\section{Acknowledgements}

The authors are grateful to K. Inoue, A. Bogdanov,  Y. Togawa, M.
Mostovoy and S. Bordacs for useful discussions. This work was funded
by JSPS Core-to-Core Program, Advanced Research Networks (Japan).
This work was supported by the Hungarian Research Fund OTKA K
108918.


%

\end{document}